\documentclass[aps,prl,twocolumn,superscriptaddress]{revtex4-2}
\usepackage{graphicx}
\usepackage{amsmath,amssymb,color}
\usepackage{}

\begin{document}



\title{Decoherence Induced Exceptional Points in a Dissipative Superconducting Qubit}

\author{Weijian Chen}\email[]{wchen34@wustl.edu}
\affiliation{Department of Physics, Washington University, St. Louis, MO, USA, 63130}
\affiliation{Center for Quantum Sensors, Washington University, St. Louis, MO, USA, 63130}

\author{Maryam Abbasi}
\affiliation{Department of Physics, Washington University, St. Louis, MO, USA, 63130}

\author{Byung Ha}
\affiliation{Department of Physics, Washington University, St. Louis, MO, USA, 63130}

\author{Serra Erdamar}
\affiliation{Department of Physics, Washington University, St. Louis, MO, USA, 63130}

\author{Yogesh N. Joglekar}
\affiliation{Department of Physics, Indiana University-Purdue University Indianapolis, Indianapolis, IN, USA, 46202}

\author{Kater W. Murch}\email[]{murch@physics.wustl.edu}
\affiliation{Department of Physics, Washington University, St. Louis, MO, USA, 63130}
\affiliation{Center for Quantum Sensors, Washington University, St. Louis, MO, USA, 63130}


\date{\today}

\begin{abstract}

Open quantum systems interacting with an environment exhibit dynamics described by the combination of dissipation and coherent Hamiltonian evolution. 
Taken together, these effects are captured by a  Liouvillian superoperator. The degeneracies of the (generically non-Hermitian) Liouvillian are exceptional points, which are associated with critical dynamics as the system approaches steady state. We use a superconducting transmon circuit coupled to an engineered environment to observe two different types of Liouvillian exceptional points that arise either from the interplay of energy loss and decoherence or purely due to decoherence. By dynamically tuning the Liouvillian superoperators in real time we observe a non-Hermiticity-induced chiral state transfer. Our study motivates a new look at open quantum system dynamics from the vantage of Liouvillian exceptional points, enabling applications of non-Hermitian dynamics in the understanding and control of open quantum systems.

\end{abstract}

\maketitle

Exceptional points degeneracies (EPs) have been extensively studied in classical dissipative systems with energy or particle loss where the dynamics are governed by effective non-Hermitian Hamiltonians \cite{Miri2019,Ozdemir2019}. Recently, there is growing interest to harness non-Hermiticities for quantum applications ranging from sensing \cite{lau2018,mcdonald2020,Yu2020} to state control \cite{liu2020, abbasi2021}. Various approaches have been used to implement non-Hermitian Hamiltonians in quantum systems such as introducing a mode-selective loss \cite{Xiao2017, Klauck2019}, embedding the desired non-Hermitian Hamiltonian into a larger Hermitian system \cite{Wu2019,liu2020}, or removing quantum jumps from the evolution of an open quantum system through postselection \cite{Naghiloo2019,abbasi2021}. However, despite its essential role in quantum systems, decoherence has not been a focus of these studies. Indeed, decoherence and its effects cannot be captured by an effective non-Hermitian Hamiltonian formalism.  Liouvillian superoperators have been proposed to take account of both the energy loss and decoherence, capturing the full dynamics of a decohering non-Hermitian system  \cite{Hatano2019,Minganti2019,Minganti2020,Arkhipov2020_02,Chen2021}. In the Liouvillian formalism, the dissipative effects are captured by Lindblad dissipators, whose effects come in two parts: one is a coherent nonunitary evolution (i.e., energy or particle loss) and the other is quantum jumps between the energy levels that lead to decoherence \cite{Dalibard1992,Plenio1998}.  
This formalism provides a critical examination when generalizing phenomena and applications observed in classical systems to quantum systems such as EP sensors \cite{wiersig2014,Chen2017,Wiersig2020}. The Liouvillian superoperators also exhibit EPs, but these LEPs and their properties have not yet been experimentally observed.  In this Letter, we study the transient dynamics of a dissipative superconducting qubit as it evolves toward its steady state. We observe LEPs that arise from the interplay of energy loss and decoherence. By dynamically tuning the Liouvillian superoperator in real time we observe a non-Hermiticity-induced chiral state transfer. 
Further, by expanding the dimension of the Hilbert space from two to three, we construct a subspace where the non-Hermiticity is purely due to decoherence. Our study shows the rich features and potential applications of non-Hermitian physics and EPs beyond the Hamiltonian formalism, further enriching applications of open quantum systems in quantum information technology. 


\begin{figure*}[htbp]
\centering
\includegraphics{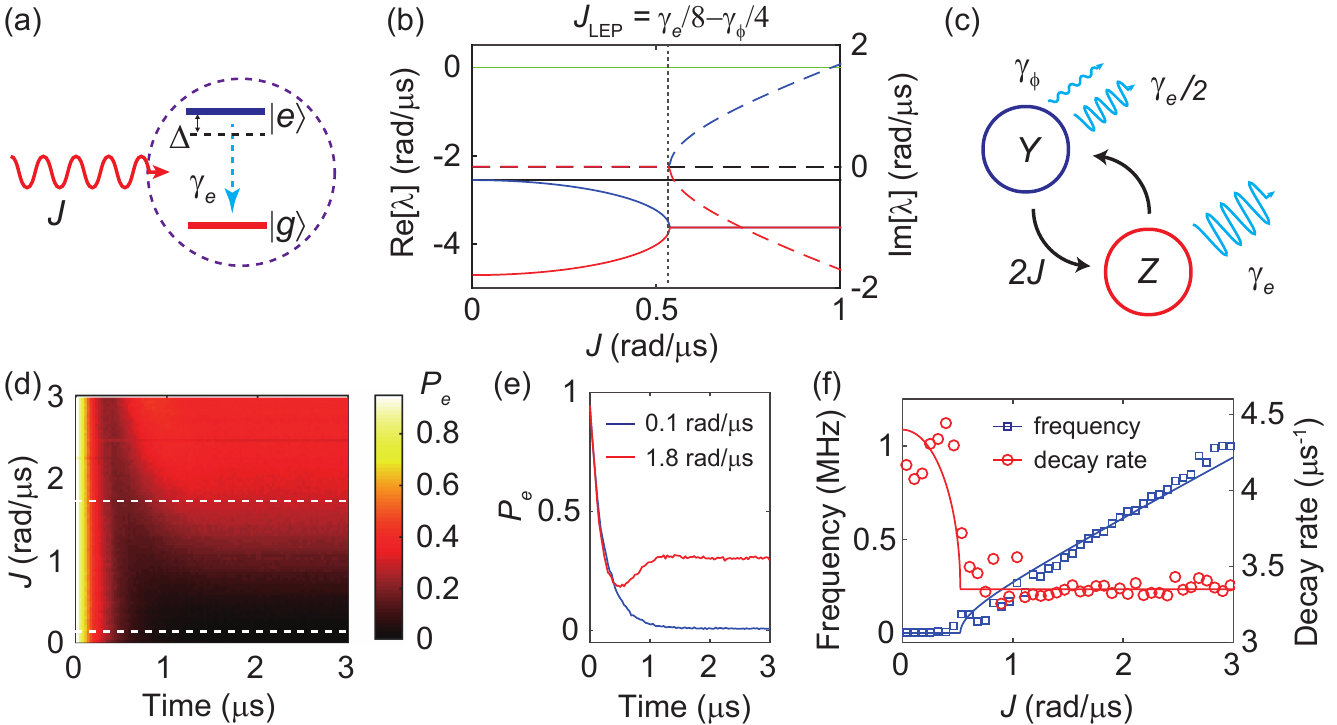}
    \caption{{\bf Liouvillian EP in the dynamics of a driven dissipative qubit.} (a) Schematic of the system; $\gamma_e$ denotes the spontaneous emission rate of the $|e\rangle$ level, and $J$ denotes the coupling rate from an applied drive with frequency detuning $\Delta$ relative to the $|g\rangle$--$|e\rangle$ transition.
    (b) Real (solid curves) and imaginary (dashed curves) parts of the Liouvillian spectra when $\Delta=0$; the LEP is indicated with a vertical dashed line. (c) The coupling between the Pauli expectation values $y$ and $z$ with different losses can be viewed in terms of a two-mode system ($Y,Z$) with passive PT symmetry and EPs. (d) Population dynamics versus evolution time at different $J$ values with the qubit initialized at the $|e\rangle$ state. Two examples (marked by the white dashed lines) of population evolution at $J = 0.1\,\mathrm{rad\,\mu \mathrm{s}^{-1}}$ (blue curve) and $J = 1.8\,\mathrm{rad\,\mu \mathrm{s}^{-1}}$ (red curve) are shown in (e). 
    (f) Oscillation frequency (blue squares) and decay rate (red circles) at different drive amplitudes; the transition marks a LEP. The solid curves are calculated from the Liouvillian spectra, where the dissipation rates $\gamma_e = 4.4\,\mu \mathrm{s}^{-1}$ and $\gamma_\phi = 0.1\,\mu \mathrm{s}^{-1}$ are used.
    }
\label{fig1}
\end{figure*}



The dynamics of a driven dissipative two-level system [Fig. \ref{fig1}(a)] can be described by a Lindblad master equation:
\begin{equation}
\dot{\rho} = -i [H_c, \rho] + \sum_{k=e,\phi} [L_k \rho L_k^\dag - \frac{1}{2} \{L_k^\dag L_k, \rho\}] \equiv \mathcal{L} \rho,    
\end{equation}
where $\rho$ denotes the density operator, and $L_{e,\phi}$ are the jump operators, defined as $L_e = \sqrt{\gamma_e} |g\rangle \langle e|$ and $L_\phi=\sqrt{\gamma_\phi/2} \sigma_z$, describing spontaneous emission from level $|e\rangle$ to level $|g\rangle$ at a rate $\gamma_e$ and pure dephasing at a rate $\gamma_\phi$, respectively. $H_\mathrm{c} = J (\vert g \rangle \langle e \vert + \vert e \rangle \langle g \vert) + \Delta/2 (\vert g \rangle \langle g \vert - \vert e \rangle \langle e \vert)$, characterizes coupling between two levels by a drive with the frequency detuning $\Delta$ relative to the $|g\rangle$--$|e\rangle$ transition at a rate $J$. The dynamics can be fully captured by a Liouvillian superoperator $\mathcal{L}$. Given a Hilbert space of dimension $N$, the Liouvillian approach is based on  representation of the system state as a density operator and the corresponding Liouville space has a dimension of $N^2$. The eigenvalues of the Liouvillian superoperator are provided in Fig.~\ref{fig1}(b) for $\Delta=0$, and there is a second-order LEP degeneracy at $J_\mathrm{EP}=\gamma_e/8-\gamma_\phi/4$.

Physical intuition for this LEP can be obtained by recasting the Lindblad equation into a Bloch equation for the expectation values of the Pauli operators $\{x, y, z\} \equiv \{\langle\sigma_x\rangle, \langle\sigma_y\rangle, \langle\sigma_z\rangle \}$ \cite{Shallem2015},
\begin{equation}
\begin{pmatrix}
\dot{x} \\ \dot{y} \\ \dot{z}    
\end{pmatrix}
= 
-\begin{pmatrix}
\frac{\gamma_e}{2}+\gamma_\phi & \Delta & 0 \\
-\Delta & \frac{\gamma_e}{2} +\gamma_\phi & 2J \\
0 & -2J & \gamma_e
\end{pmatrix}
\begin{pmatrix}
x \\ y \\ z    
\end{pmatrix}
+
\begin{pmatrix}
0 \\ 0 \\ \gamma_e    
\end{pmatrix}.
\end{equation}
The $y$ and $z$ components are coupled, yet exhibit different losses, yielding effectively a passive parity-time (PT) symmetric system [Fig.~\ref{fig1}(c)]. 
The $z$ component exhibits a loss of excitation (energy), whereas the $y$ component exhibits decoherence from both spontaneous emission and pure dephasing. There is one LEP for $\Delta=0$ except when $\gamma_e = 2 \gamma_\phi$, where there is no longer any loss contrast.  

In the experiment, we use the lowest two energy levels ($\vert g \rangle$, $\vert e \rangle$) of a transmon superconducting circuit \cite{Koch2007}. The transmon is dispersively coupled to a three-dimensional microwave cavity, leading to a state-dependent cavity resonance frequency. High fidelity, single-shot readout of the transmon state can be realized by probing the cavity with a weak microwave signal and detecting its phase shift \cite{Wallraff2005}. Further, we shape the density of states of the electromagnetic field which allows us to adjust the dissipation rate of the energy level $|e\rangle$ \cite{Naghiloo2019}. In this study, we set $\gamma_e \approx 4.5\,\mu \mathrm{s}^{-1}$, much greater than the pure dephasing rate $\gamma_\phi \approx 0.2\,\mu \mathrm{s}^{-1}$ so that there is a large loss contrast between $y$ and $z$.

To experimentally identify the LEP, we study the transient dynamics of the qubit to its steady state. We initialize the qubit in  the $|e\rangle$ state and then apply a resonant microwave drive to induce a coupling at rate $J$. Figure~\ref{fig1}(d) displays the measured evolution of the $|e\rangle$ state population for different $J$. We observe a transition from exponential decay to exponentially damped oscillation as the coupling rate is increased.   Examples of the evolution at two different drive amplitudes above and below the LEP are shown in Fig.~\ref{fig1}(e). A classical analogy of this observation is a damped harmonic oscillator, where a second-order EP (corresponding to critical damping) marks the transition from an overdamped to an underdamped regime. 
The results are processed by fitting to a decaying sine wave to determine the oscillation frequency and decay rate [Fig.~\ref{fig1}(f)], which show a transition at $J \simeq \gamma_e/8$, in agreement with the Liouvillian eigenvalues.

\begin{figure*}[htbp]
\centering
\includegraphics[width=6 in]{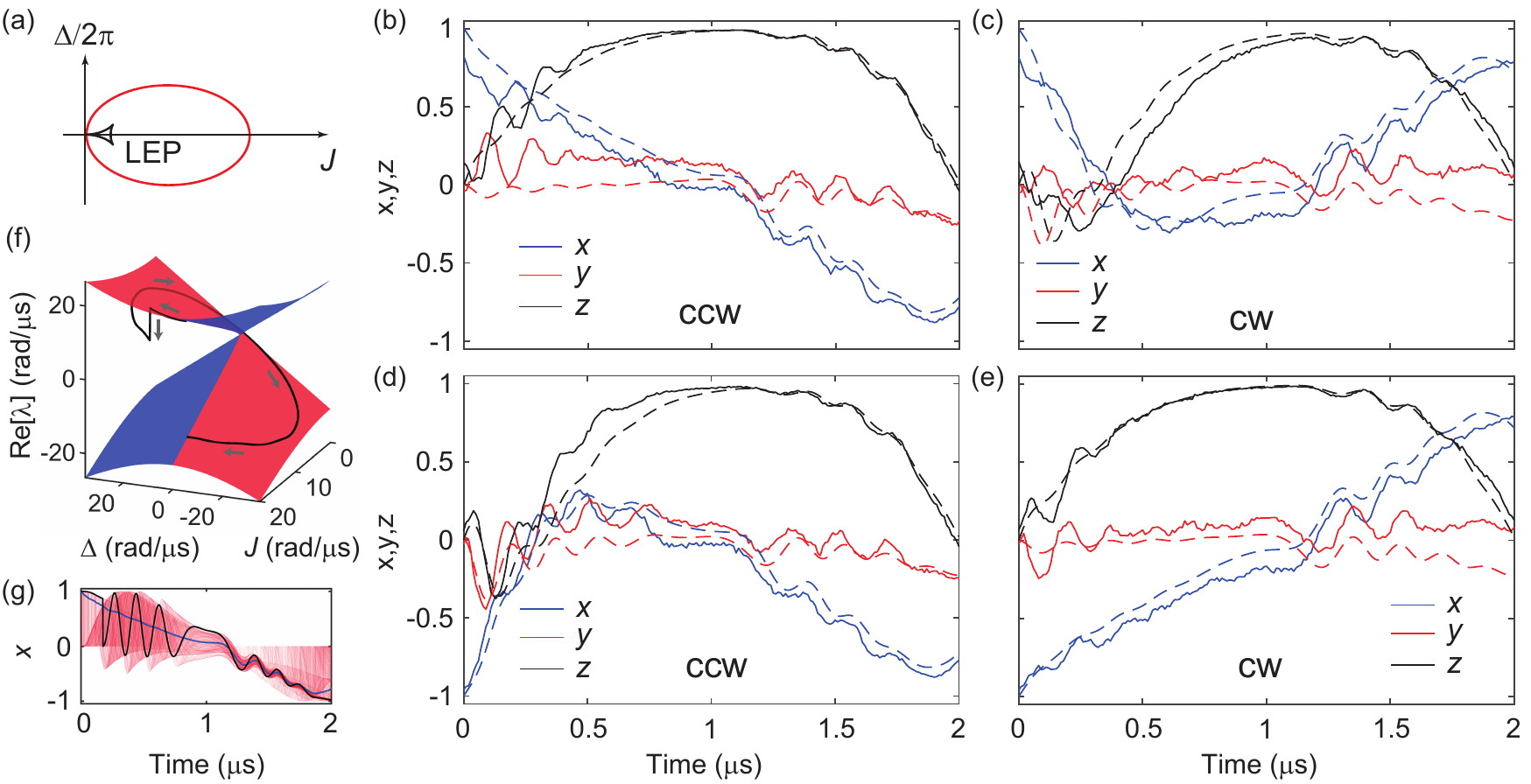}
    \caption{{\bf Dynamical encircling of the Liouvillian EP structure.} (a) Schematic of the parameter path in the parameter space $(J,\Delta)$, where the LEP structure is magnified by 20 times for clarity. (b-e) Evolution of the Bloch components under two different initial states ($|\!\pm\!x\rangle$) and two encircling directions (cw, ccw). The solid curves are the experimental results, and the dashed curves are the theoretical calculations from Lindblad master equation, with $\gamma_e = 4.6\, \mathrm{\mu \mathrm{s}^{-1}}$ and $\gamma_\phi = 0.2\, \mathrm{\mu \mathrm{s}^{-1}}$. (f) Illustration of one quantum trajectory (marked by grey arrows) on the Riemann surface, where there is one quantum jump. This trajectory is one example (black curve) from $1000$ simulated trajectories (red curves) shown in (g). The average of the trajectories is the solution of the Lindblad equation (blue curve).}
\label{fig2}
\end{figure*}

One signature of Hamiltonian based EPs is the chiral state transfer that occurs when the Hamiltonian parameters are tuned to encircle an EP. As a result of the topological structure of the Riemann manifold that describes the system's complex energy, one state will map to the other after the encirclement. Relative gain or loss along different paths results in chiral state or population transfer \cite{Xu2016,Doppler2016,Choi2017,liu2020,abbasi2021}. This process has also been shown to induce a chiral geometric phase on quantum states \cite{abbasi2021}. Here, we investigate whether these population features persist when encircling the LEPs in the parameter space $(J,\Delta)$. 

For non-zero $\Delta$, the Liouvillian exhibits second-order LEP lines and two third-order LEPs, forming a small ``LEP structure" very near the LEP for $\Delta=0$ \cite{supp}. We now investigate the effects of dynamically tuning the Liouvillian parameters to encircle this LEP structure.   We choose a closed parameter path defined as $J(t) = 16 \cos^2(\pi t/T)\, \mathrm{rad\,\mu \mathrm{s}^{-1}}$ and $\Delta (t) = \pm 10\pi \sin (2\pi t/T)\, \mathrm{rad\,\mu \mathrm{s}^{-1}}$, where $T =2~\mu$s is the loop period, and $``+"$ and $``-"$ correspond to counter-clockwise (ccw) and clockwise (cw) encircling directions, respectively [Fig. \ref{fig2}(a)]. We choose the initial state $|\!+\!x\rangle$ at $t=0$. The results of quantum state tomography at different points along the parameter path for both cw and ccw directions are shown in Fig.~\ref{fig2}(b, c). While for the ccw direction, the initial state is transferred to a state close to $|\!-\!x\rangle$, for the cw direction, the final state remains approximately at $|\!+\!x\rangle$.  Similar observations also apply to the case with the initial state $|\!-\!x\rangle$ [Fig.~\ref{fig2}(d,e)]. 

This chiral behavior can be understood from a quantum trajectory picture. The qubit evolution can be described by a non-Hermitian Hamiltonian evolution that is interrupted by randomly occurring quantum jumps. The non-Hermitian Hamiltonian evolution pertains to the Riemann structure displayed in Fig.~\ref{fig2}(f), which would induce a state transfer upon one encircling. Figure~\ref{fig2}(f) displays one such trajectory where a quantum jump occurs. The initial state is $|\!+\!x\rangle$, and a jump to $|g\rangle$ occurs shortly after the beginning of the parameter sweep (at $t\simeq 0.2~\mu \mathrm{s}$). This state continues to evolve under the time dependent Hamiltonian. 
Remarkably, at the end of the parameter sweep, the final state is near $|\!-\!x\rangle$. An ensemble of such trajectories is shown in Fig.~\ref{fig2}(g). This chirality of state transfer originates from the directionality of the quantum jumps which favors the ground state and therefore disappears in the Hermitian limit (see the Supplementary Materials \cite{supp}).

We highlight several aspects that are different from previous studies of encircling EPs based on non-Hermitian Hamiltonians. First, in previous studies, the initial state is usually chosen to be an eigenstate. However, here the initial states $|\!\pm\!x\rangle$ do not directly correspond to the eigenstates of the Liouvillian superoperators; instead, they are approximately a superposition of two Liouvillian eigenstates, one of which corresponds to the steady state, the other an unphysical state \cite{Minganti2019, supp}. Second, the evolution is trace preserving; in contrast, for the evolution governed by non-Hermitian Hamiltonian \cite{abbasi2021}, the state norm decreases with time, and a state re-normalization at each time step is then required. Third, the quantum state is mixed due to the decoherence, which will limit the practical applications of this chiral state transfer protocol. As we show in the Supplementary Materials \cite{supp}, the decoherence effects can be minimized by optimizing the driving conditions while maintaining the chiral behavior.

\begin{figure}[htbp]
\centering
\includegraphics{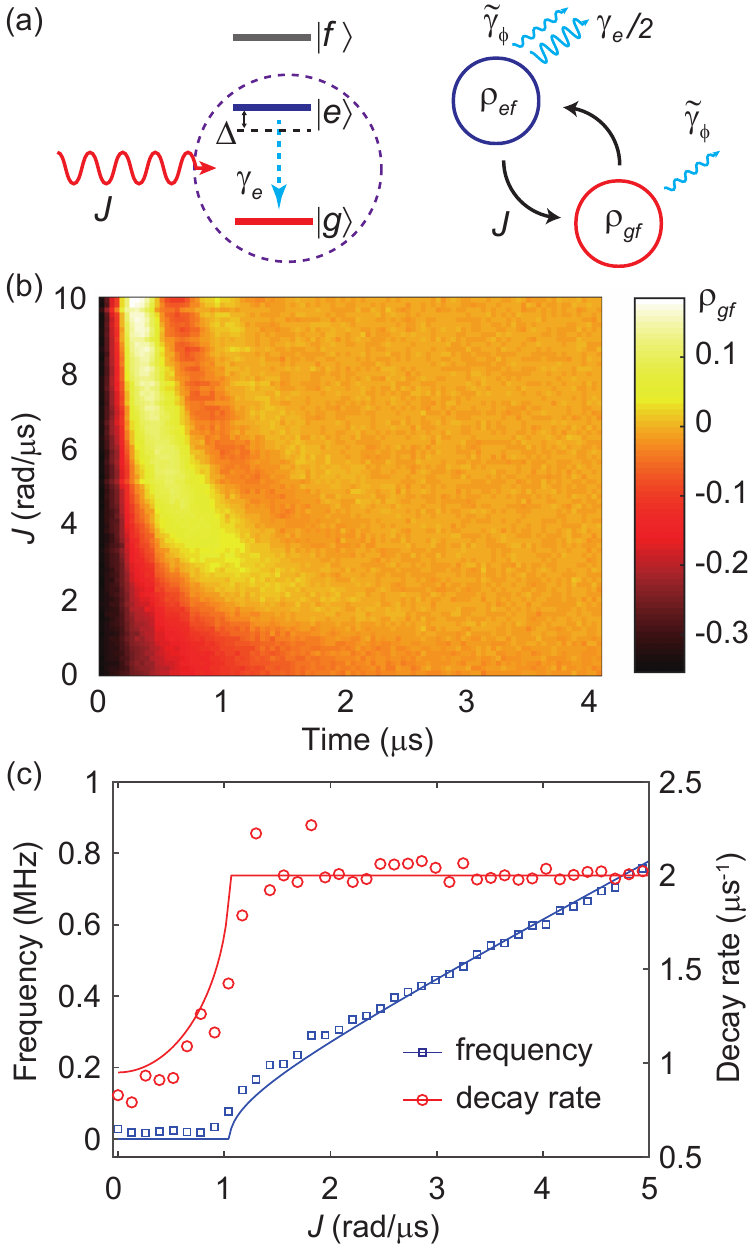}
    \caption{{\bf Decoherence induced Liouvillian EP.} (a) Schematic of a driven dissipative qutrit. The physical origin of this LEP arises from the coupling between two coherences of the density matrix $\rho_{gf}$ and $\rho_{ef}$ that experience unbalanced losses. The two coherences have the same loss $\tilde{\gamma}_\phi$ from the dissipations of the reference level $|f\rangle$, but $\rho_{ef}$ has additional loss from $\gamma_e$.  
    (b) The measured coherence $\rho_{gf}$ versus evolution time at different $J$ values. (c) Oscillation frequency (blue squares) and decay rate (red circles) at different drive amplitudes, where the transition marks a LEP. The solid curves are calculated from the Liouvillian spectra. Parameters used are $\gamma_e = 4.2\ \mu \mathrm{s}^{-1}$, $\gamma_\phi = 0.2\ \mu \mathrm{s}^{-1}$, $\gamma_f = 0.3\ \mu \mathrm{s}^{-1}$, and an additional overall loss $0.75\ \mu \mathrm{s}^{-1}$ is added to account for additional decoherence of the $|f\rangle$ state. }
\label{fig4}
\end{figure}

So far, we have only focused on the lowest two levels of the transmon circuit. By including a higher energy level (i.e., the $|f\rangle$ level, with spontaneous decay rate $\gamma_f \ll \gamma_e$) as a coherence reference [Fig. \ref{fig4}(a)], we discover a second type of LEP that is fully induced by decoherence and has no energy loss involved. 
The corresponding Liouville space then has a dimension of $9$, and the Liouvillian spectra are provided in the Supplementary Materials \cite{supp}. 
The decoherence induced LEP results from the coupling between two coherences $\rho_{gf}$ and $\rho_{ef}$ of this qutrit and occurs at $J = \gamma_e/4$ [Fig.~\ref{fig4}(a)]. The dissipation of the $|e\rangle$ level leads to the loss of the coherence $\rho_{ef}$, but not the coherence $\rho_{gf}$. As with energy loss induced EPs, the interplay between coupling and decoherence yields this LEP.  

To observe this LEP transition, we initialize the circuit in the state $(|g\rangle-|f\rangle)/\sqrt{2}$ and then apply a resonant drive with variable duration to $\{|g\rangle,|e\rangle\}$ transition, followed by a tomography pulse to determine $\rho_{gf}$ \cite{Thew2002,Bianchetti2010}. As displayed in Fig.~\ref{fig4}(b), for large $J$, we observe damped oscillations in $\rho_{gf}$, yet for $J < 1\ \mathrm{rad}/\mu\mathrm{s}$ the oscillations are replaced with exponential decay. We quantify this transition by fitting the $\rho_{gf}$ evolution to a damped sine wave, extracting the frequency and decay rate as displayed in Fig.~\ref{fig4}(c). 
We note that this decoherence induced LEP is non-local in the sense that it only relies on initial coherence between the $|f\rangle$ and $|g\rangle$ states, but no further coupling between the $\{|f\rangle\}$ and $\{|g\rangle,|e\rangle\}$ manifolds. Therefore, we expect such decoherence induced LEPs to play a critical role in how many-body correlations decay due to local operations and sources of dissipation.


Our study has revealed and quantified two new types of EPs occurring in single dissipative quantum systems. In contrast to prior work, these LEPs do not rely on postselection to induce non-Hermitian dynamics but instead are evident in the transient dynamics of an open quantum system as it approaches steady state. Because the Liouvillian formalism applies to all Markovian dissipative interactions it also encompasses the effects of quantum measurement \cite{decastro2019}.  For instance, the quantum Zeno effect pertains to the competition between coherent coupling and the dissipative effects of measurement \cite{Presi1996,Kaku2015}. The transition from a Zeno pinning regime can naturally be treated in the context of LEPs introduced here \cite{Kumar2020,chentao2021,li2020}. Our study therefore motivates a new look at open quantum system dynamics from the vantage of Liouvillian exceptional points, enabling applications of non-Hermitian dynamics in Floquet physics \cite{gunderson2021}, quantum steering \cite{kumar2021}, state transfer \cite{Pick2019,kumar2021_2}, measurement induced dynamics \cite{haco18,wang21}, and quantum thermal engines \cite{khandelwal2021}.



We acknowledge K. Snizhko and P. Kumar for inspiring discussions. This research was supported by NSF Grant No. PHY-1752844 (CAREER), AFOSR MURI Grant No. FA9550-21-1-0202, ONR Grant No. N00014-21-1-2630, and the Institute of Materials Science and Engineering at Washington University.


%
%

\widetext
\begin{center}
	\textbf{\large Supplemental Information for ``Decoherence Induced Exceptional Points in a Dissipative Superconducting Qubit''}
\end{center}

In these supplementary materials, we describe our experimental setup, provide the spectra of the Liouvillian superoperators in detail and more experimental results.
	
\section{A. Experimental Setup}

The experiment consists of a transmon superconducting circuit embedded in a three-dimensional copper cavity. The transmon is fabricated in a SQUID geometry, allowing tuning of the energy levels. We set $\omega_{ge}/2\pi=5.71\,\mathrm{GHz}$ and $\omega_{ef}/2\pi=5.41\,\mathrm{GHz}$, with the corresponding charging energy $E_c/h = 270\,\mathrm{MHz}$ and the Josephson energy $E_J/h = 16.6\,\mathrm{GHz}$. The strongly coupled port of the cavity is connected via a microwave cable to an impedance mismatch, which shapes the density of states in the transmission line, yielding a frequency dependence of the transmon decay rates.

The microwave cavity (with the dressed resonance frequency $\omega_c/2\pi = 6.684\,\mathrm{GHz}$ and decay rate $\kappa_c/2\pi = 5\,\mathrm{MHz}$) allows us to perform high-fidelity single-shot readout of the qutrit states. The dispersive coupling rates are given by $\chi_e/2\pi = -2\,\mathrm{MHz}$ and $\chi_f/2\pi = -11\,\mathrm{MHz}$.  The readout fidelity for $|g\rangle$, $|e\rangle$, $|f\rangle$ are about $77\%$, $78\%$, $98\%$, respectively. 
In quantum state tomography of $\rho_{gf}$, a $\pi$ pulse (resonant to $|g\rangle$--$|e\rangle$ transition) is first used to flip qubit state in the $\{|g\rangle,|e\rangle\}$ submanifold, and then a $\pi/2$ pulse (resonant to $|e\rangle$--$|f\rangle$ transition) is used to rotate the submanifold about X and Y axes (for $\mathrm{Re}[\rho_{gf}]$ and $\mathrm{Im}[\rho_{gf}]$), followed by readout in the energy eigenbasis. The duration of $\pi$ pulses used in the sequence are $\tau_{ge}=34\,\mathrm{ns}$ and $\tau_{ef}=30\,\mathrm{ns}$, respectively.  

\section{B. Spectra of the Liouvillian superoperator}

\begin{figure*}[htbp]
\centering
\includegraphics{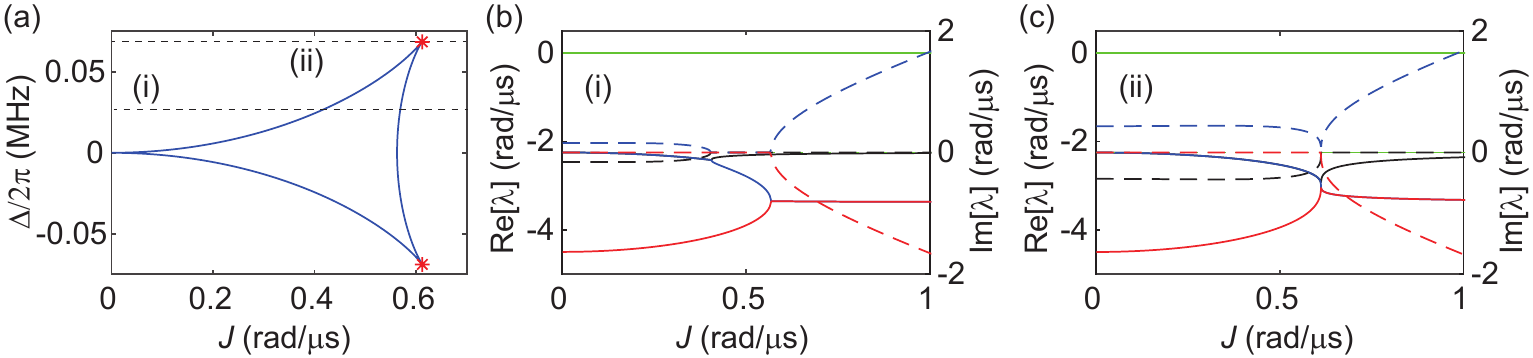}
    \caption{(a) Distribution of the LEPs in the parameter space $(J, \Delta)$, consisting of three (second-order) exceptional lines and two third-order EPs (marked by stars). The real (solid curves) and imaginary (dashed curves) parts of the eigenvalues along the two black dashed lines are shown in (b,c). Parameters used are: $\gamma_e = 4.5\ \mu \mathrm{s}^{-1}$, $\gamma_\phi = 0$, $\Delta/2\pi = 0.025\ \mathrm{MHz}$ (b) and $0.069\ \mathrm{MHz}$ (c).}
\label{EPlines}
\end{figure*}

The dynamics of a dissipative two-level quantum system in our study is described by a Lindblad master equation 
\begin{equation}
\frac{\partial \rho}{\partial t} = -i [H_\mathrm{c}, \rho] + \sum_{k=e,\phi} [ L_k \rho L_k^\dag - \frac{1}{2} \{L_k^\dag L_k, \rho\} ] \equiv \mathcal{L}\rho,
\end{equation}
where $\rho$ denotes a $2 \times 2$ density operator. The jump operators $L_e = \sqrt{\gamma_e} |g\rangle \langle e|$ and $L_\phi = \sqrt{\gamma_\phi/2} \sigma_z$ describe the energy decay from $|e \rangle$ to $|g \rangle$ and the pure dephasing, respectively. In the rotating frame $H_\mathrm{c} = J (|g \rangle \langle e| + |e \rangle \langle g|) + \Delta/2 (|g \rangle \langle g| - |e \rangle \langle e|)$, where $\Delta$ is the frequency detuning (relative to the $|g\rangle$--$|e\rangle$ transition) of the microwave drive that couples the states at rate $J$. To study the Liouvilian spectra and LEPs, we first represent the Liouvillian superoperator in a matrix form \cite{Minganti2019,Minganti2020}, given by
\begin{equation}
    \mathcal{L}^\mathrm{matrix} = -i(H_c\otimes I - I\otimes H_c^\mathrm{T}) 
    + \sum_{k=e,\phi} [ L_k\otimes L_k^{\ast} 
    - \frac{L_k^{\dagger} L_k\otimes I}{2} - \frac{I \otimes L_k^\mathrm{T} L_k^{\ast}}{2} ],
\end{equation}
where $\otimes$ represents Kronecker product operation, $\mathrm{T}$ represents the transpose, and $\ast$ represents the complex conjugate. 
The matrix form of the Liovillian superoperator for the qubit is given by,
\begin{equation}
\mathcal{L}_\mathrm{qubit}^\mathrm{matrix}
= 
\begin{pmatrix}
0 & iJ & -iJ & \gamma_e \\
iJ & -i\Delta-\gamma_e/2 - \gamma_\phi & 0 & -iJ \\
-iJ & 0 & i\Delta-\gamma_e/2 - \gamma_\phi & iJ \\
0 & -iJ & iJ & -\gamma_e
\end{pmatrix}.
\end{equation}
The Liouvillian matrix exhibits a triangle-shaped structure of LEPs in the parameter space $(J, \Delta)$, consisting of three exception lines (of the second-order LEPs) and two third-order LEPs (Fig.~\ref{EPlines}) \cite{Noh2010,Shallem2015}.

When $\Delta=0$, the eigenvectors of the Liouvillian superoperators written in matrix form are given by, 
\begin{equation}
    \rho_0 \propto \frac{1}{\gamma_e^2 + 8J^2}
    \begin{pmatrix}
    \gamma_e^2 + 4J^2 & 2i\gamma_e J \\
    -2i\gamma_e J & 4 J^2
    \end{pmatrix},
\end{equation}
\begin{equation}
    \rho_1 \propto
    \begin{pmatrix}
    0 & 1 \\
    1 & 0
    \end{pmatrix},
\end{equation}
\begin{equation}
    \rho_{2,3} \propto
    \begin{pmatrix}
    -\gamma_e \mp \sqrt{\gamma_e^2 - 64J^2}  & 8iJ \\
    -8iJ & \gamma_e \pm \sqrt{\gamma_e^2 - 64J^2}
    \end{pmatrix},
\end{equation}
where $\gamma_\phi=0$. $\rho_0$ is the eigenvector with zero eigenvalue and corresponds to the steady state. $\rho_1$ is only related to the $X$ Bloch component and doesn't participate in the LEP transition (in $\Delta=0$ case). $\rho_{2,3}$ are related with a second-order LEP (denoted as type-I LEP) and only contain $Y$ and $Z$ Bloch components. Therefore, to observe the LEP, the initial state of the qubit cannot be the steady state. 

\begin{figure*}[htbp]
\centering
\includegraphics{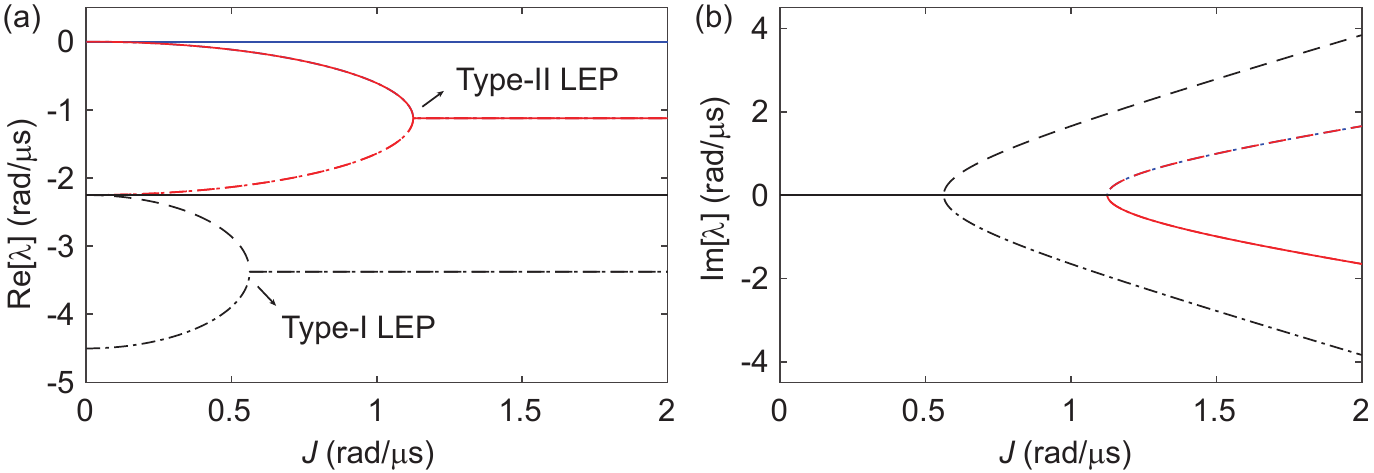}
    \caption{Real (a) and imaginary (b) parts of the eigenvalues of the Liouvillian superoperator for the qutrit. Two types of LEPs can be observed. Parameters used are: $\gamma_e = 4.5\ \mu \mathrm{s}^{-1}$, $\gamma_\phi = 0$, $\Delta = 0$.}
\label{99spectrum}
\end{figure*}

\begin{figure*}[htbp]
\centering
\includegraphics{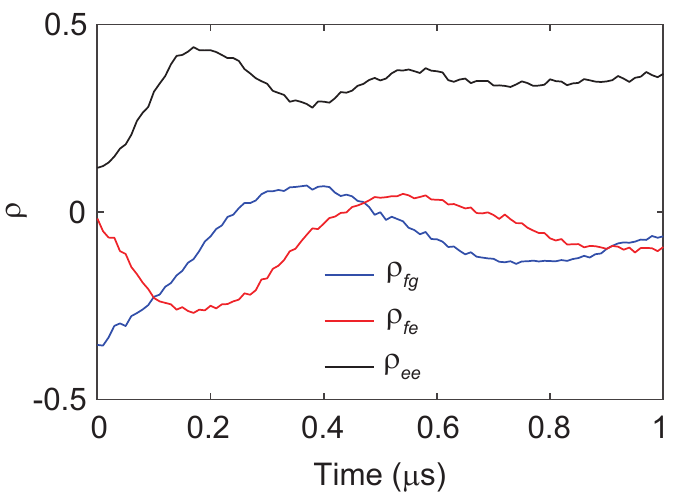}
    \caption{Transient dynamics of $\rho_{ee}$ and two coherence terms $\rho_{fg}$ and $\rho_{fe}$ under a resonant drive in the $\{|g\rangle,|e\rangle\}$ submanifold with $J = 8\ \mathrm{rad}\, \mu \mathrm{s}^{-1}$. The qutrit is prepared at $(|g\rangle-|f\rangle)/\sqrt{2}$ at $t=0$.}
\label{coherence_coup}
\end{figure*}

The above method can be directly extended to a qutrit. For the qutrit with spontaneous decay from the $|e\rangle$ state at rate $\gamma_e$, the Liouvillian at $\Delta=0$ takes the form of a $9\times 9$ matrix,
\begin{equation}
\mathcal{L}_\mathrm{qutrit}^\mathrm{matrix} 
= 
\begin{pmatrix}
0 & iJ & 0 & -iJ & \gamma_e & 0 & 0 & 0 & 0 \\
iJ & -\gamma_e/2 & 0 & 0 & -iJ & 0 & 0 & 0 & 0 \\
0 & 0 & 0 & 0 & 0 & -iJ & 0 & 0 & 0 \\
-iJ & 0 & 0 & -\gamma_e/2 & iJ & 0 & 0 & 0 & 0 \\
0 & -iJ & 0 & iJ & -\gamma_e & 0 & 0 & 0 & 0 \\
0 & 0 & -iJ & 0 & 0 & -\gamma_e/2 & 0 & 0 & 0 \\
0 & 0 & 0 & 0 & 0 & 0 & \boldsymbol{0} & \boldsymbol{iJ} & 0 \\
0 & 0 & 0 & 0 & 0 & 0 & \boldsymbol{iJ} & \boldsymbol{-\gamma_e/2} & 0 \\
0 & 0 & 0 & 0 & 0 & 0 & 0 & 0 & 0 \\
\end{pmatrix}. \label{eq:99lse}
\end{equation} 
The bold terms in Eq.~(\ref{eq:99lse}) correspond to the decoherence induced LEP (denoted as type-II LEP) discussed in the main text. Figure~\ref{99spectrum} shows the eigenvalues of the Liouvillian matrix of the qutrit at $\Delta=0$, where both types of LEPs can be observed.

Figure~\ref{coherence_coup} shows one example of coupling between the two coherent terms $\rho_{fg}$ and $\rho_{fe}$ as well as the evolution of $\rho_{ee}$ under a constant drive $J$. 
Under the same drive, $\rho_{ee}$ that is associated with the type-I LEPs oscillates in a frequency about twice of that of the coherent terms, which is consistent with the theoretical results shown in Fig.~\ref{99spectrum}(b).

\section{C. Slow parameter driving at the Hermitian limit}

\begin{figure*}[htbp]
\centering
\includegraphics{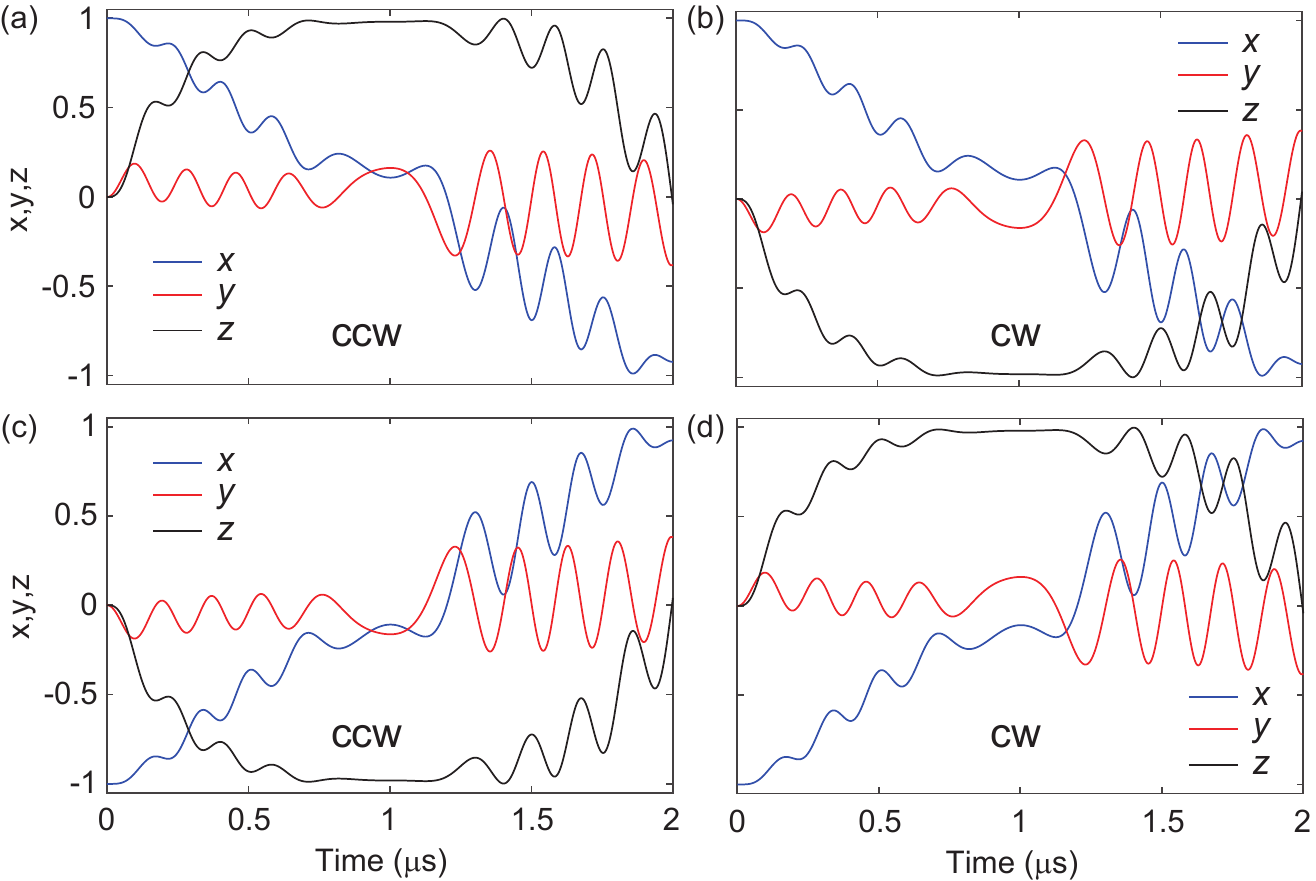}
    \caption{Time evolution of the Bloch components at two different initial states ($|\!+\!x\rangle$ or $|\!-\!x\rangle$) and two encircling directions (cw or ccw). The theoretical results are calculated by using the same parameter path and duration but assuming $\gamma_{e,\phi} = 0$.}
\label{hermitianlimit}
\end{figure*}

In the main text, we show the chiral state transfer when the system parameters are tuned in real time. This chirality  originates from the directionality of the quantum jumps which favors the ground state and therefore disappears in the Hermitian limit. Figure~\ref{hermitianlimit} shows the theoretical results of the qubit state evolution in the Hermitian limit (i.e., $\gamma_{e,\phi}=0$). Given an initial state ($|\pm x\rangle$), both encircling directions transfer the Bloch component $X$ from $\pm 1$ to about $\mp 1$.

\section{D. Chiral state transfer at different parameter paths}

\begin{figure*}[htbp]
\centering
\includegraphics{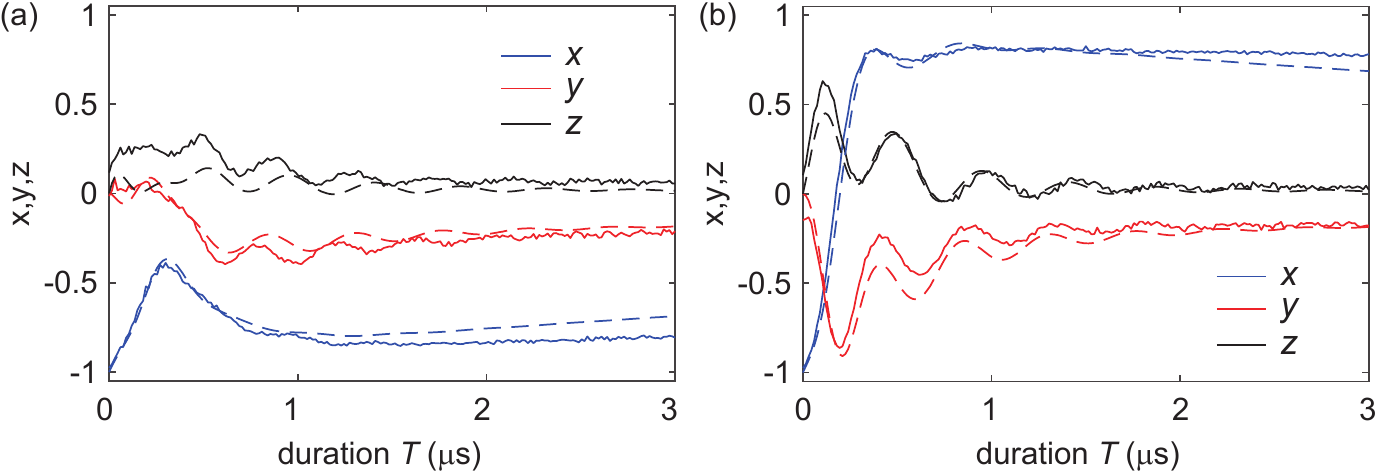}
    \caption{Quantum state tomography after one full encirclement with the ccw (a) and cw (b) directions at different duration $T$. Other path parameters are the same as those in the main text. The initial qubit state is $|\!-\!x\rangle$. The solid curves are experimental results, and the dashed curves are theoretical results from Lindblad equation.}
\label{SIfig_duration}
\end{figure*}

\begin{figure*}[htbp]
\centering
\includegraphics{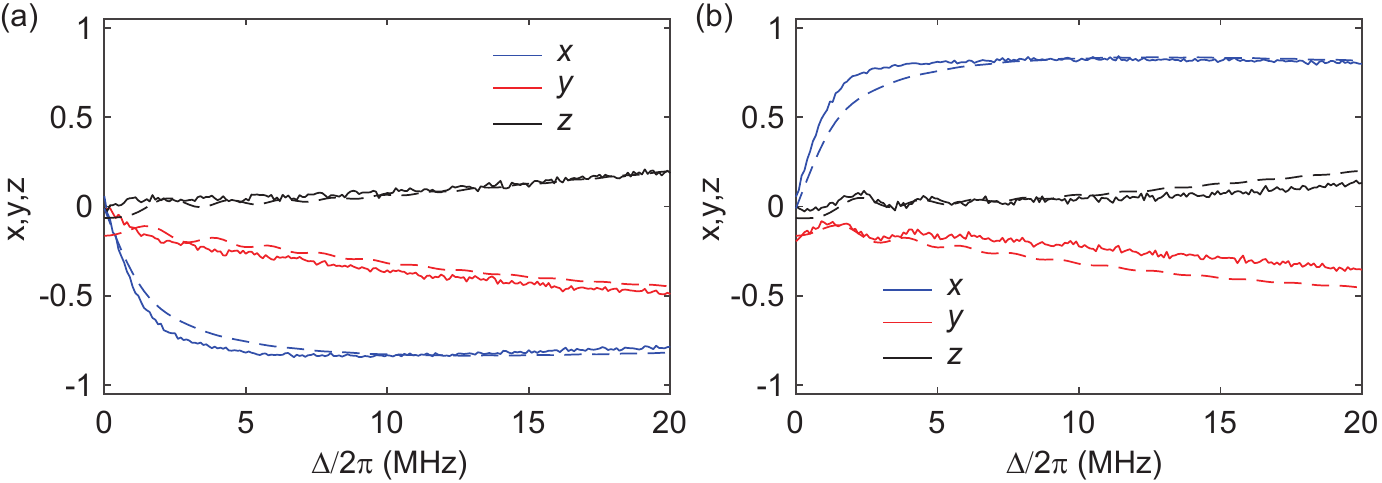}
    \caption{Quantum state tomography after one full encirclement with the ccw (a) and cw (b) directions at different maximum detuning $\Delta$. Other path parameters are the same as those in the main text. The qubit initial state is $|\!-\!x\rangle$, and the loop time is $2\,\mu\mathrm{s}$. The solid curves are experimental results, and the dashed curves are theoretical results from Lindblad equation.}
\label{SIfig_detuning}
\end{figure*}

In this section, we provide additional results of the quantum state tomography for different path parameters including duration $T$ (Fig.~\ref{SIfig_duration}) and maximum detuning $\Delta$ (Fig.~\ref{SIfig_detuning}). To quantify the chiral population transfer, we introduce two parameters: the trace distance between the final states $\rho_\mathrm{cw,ccw}$ under two encircling directions $\mathcal{C} = \frac{1}{2} \mathrm{Tr}[\sqrt{(\rho_\mathrm{cw} - \rho_\mathrm{ccw})^\dag(\rho_\mathrm{cw} - \rho_\mathrm{ccw})}]$ quantifies the chirality, and the entropy $S = -\sum_i p_i \log_2(p_i)$ quantifies the purity of the final states, where $p_i$ denotes the eigenvalues of the density operators.

\begin{figure*}[htbp]
\centering
\includegraphics{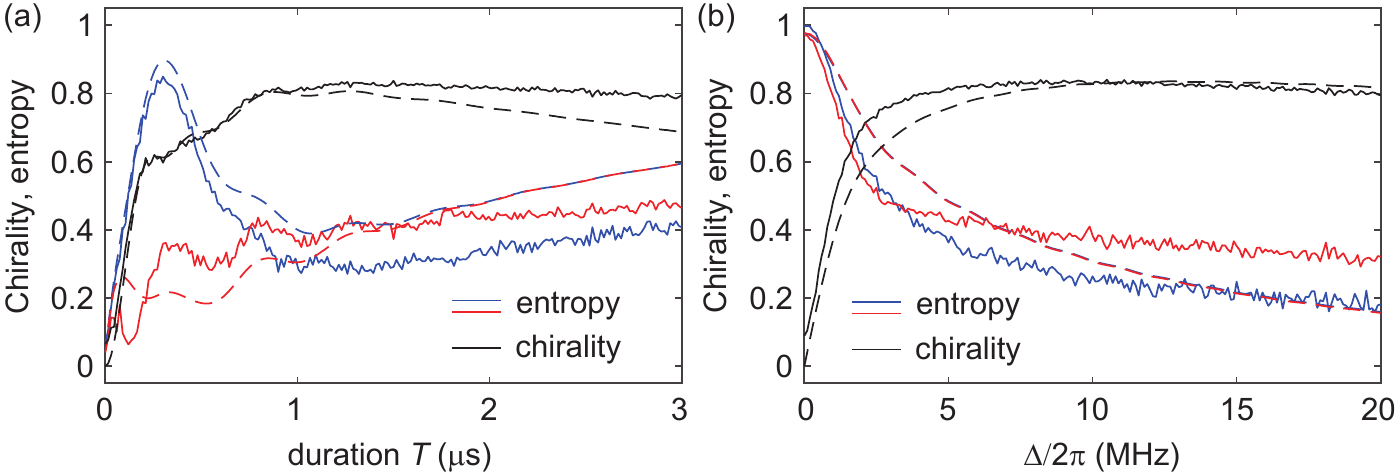}
    \caption{Dependence of entropy (blue: ccw; red: cw) and chirality (black) on the duration $T$ (a) and the maximum frequency detuning $\Delta$ (b). The maximum detuning in (a) is $2\pi \times 5\,\mathrm{MHz}$, and the loop duration in (b) is $2\,\mu \mathrm{s}$. The solid curves are experimental results, and the dashed curves are theoretical results from Lindblad equation. 
    }
\label{fig3}
\end{figure*}

Figure~\ref{SIfig_duration} shows the tomography results of the final states after one encirclement for different loop times $T$. 
There is an oscillation that can be attributed to Floquet dynamics \cite{abbasi2021}. The dependence of the chirality and entropy on the duration is shown in Fig.~\ref{fig3}(a). 
We observe maximal chirality and at the same time high purity of the final states at about $T=1\,\mathrm{\mu \mathrm{s}}$. Figure~\ref{SIfig_detuning} displays the tomography results of the final states after one encirclement for different values of the maximum detuning $\Delta$. The dependence of the chirality and entropy on $\Delta$ is provided in Fig.~\ref{fig3}(b). With increasing $\Delta$, the chirality increases, while the entropy reduces. The purity of the transferred states can be further improved by realizing 
a time-dependent dissipation rate $\gamma_e$, for instance, $\gamma_e = \gamma_{e0}[1 - \cos (2\pi t/T)]/2$, where the decoherence effect only becomes significant when there is strong relative gain and loss effect.

\end{document}